\pdfoutput=1

\documentclass[letterpaper]{article} 
\usepackage{aaai25}  
\nocopyright

\usepackage{times}  
\usepackage{helvet}  
\usepackage{courier}  
\usepackage[hyphens]{url}  
\usepackage{graphicx} 
\usepackage{natbib}  
\usepackage{caption} 
\usepackage{algorithm}
\usepackage{algorithmic}

\usepackage{booktabs}
\usepackage{caption}
\usepackage{multirow}
\usepackage{tabularx}

\usepackage{xcolor}

\usepackage{newfloat}
\usepackage{listings}
\DeclareCaptionStyle{ruled}{labelfont=normalfont,labelsep=colon,strut=off} 
\floatstyle{ruled}
\newfloat{listing}{tb}{lst}{}
\floatname{listing}{Listing}
\title{More Skin, More Likes! \\ Measuring Child Exposure and User Engagement on TikTok}
\author{
    Miriam Schirmer, Angelina Voggenreiter, Jürgen Pfeffer
}
\affiliations{
    School of Social Sciences and Technology\\
    Technical University of Munich\\

    \{firstname.lastname\}@tum.de

}

\begin{document}

\maketitle

\begin{abstract}
Sharenting, the practice of parents sharing content about their children on social media, has become increasingly common, raising concerns about children's privacy and safety online. This study investigates children's exposure on TikTok, offering a detailed examination of the platform's content and associated comments. Analyzing 432,178 comments across 5,896 videos from 115 users featuring children, we categorize content into Family, Fashion, and Sports. Our analysis highlights potential risks, such as inappropriate comments or contact offers, with a focus on appearance-based comments. Notably, 21\% of comments relate to visual appearance. Additionally, about 20\% of videos depict children in revealing clothing, such as swimwear
, attracting significantly more appearance-based comments and likes than videos featuring fully clothed children. 
These findings underscore the need for heightened awareness and protective measures to safeguard children's privacy and well-being in the digital age.

\end{abstract}

%

\section{Introduction}

New York Times journalists recently investigated over 2.1 million Instagram posts featuring children and uncovered a "marketplace of girl influencers," usually managed by their mothers \citep{nyt2024}. Often portraying young girls in exposing attire, the reporters found that these posts draw the attention of men sexually attracted to children and have shed light on the complex and potentially exploitative dynamics behind online content of minors. 
With its short-form video format and widespread popularity, TikTok has become a significant platform for self-expression, creativity, and social interaction. As of the latest available data, TikTok boasts a staggering user base, with an estimated 900 million in 2024 \citep{statista2024}. 

Despite TikTok's explicit age restrictions, prohibiting children under 13 years old from creating accounts, the platform remains a magnet for younger users. Children actively generate content that reflects their interests, talents, and daily lives, shaping the platform's ecosystem \citep{pedrouzo2023hyperconnected}. In the United States of America, for example, the largest proportion of TikTok Users (25\%) are between 10 and 19 years old \citep{YoungUsers}. Similarly, in the United Kingdom, almost one-third of 5-7-years-olds, half of 8-11-years-olds and more than two-thirds of 12-15-years-olds use TikTok \citep{ofcom2022}. 

This study explores children's exposure on TikTok. We identify and categorize the most common types of content showing minors on the platform. We analyze the comments on these videos, assessing the nature of viewer feedback and any notable aspects. Additionally, we examine specific risks for children on social media, particularly the potential for sexual exploitation on TikTok \citep{are2023flagging}. Considering previous findings that skin exposure increases user engagement \citep{kernen2021social,ramsey2018picture}, we investigate how attire and skin exposure influence likes and comments, highlighting the vulnerability of young users to exploitative behaviors, making several key contributions:

\begin{itemize}
    \item We provide one of the first studies to analyze child exposure on TikTok on a sample of 432,178 comments, showing trends in the presentation of minors and reactions to such content.
    \item We show that a substantial share of 19.57\% of the videos depict children in revealing clothing.
    \item We find that such clothing is significantly linked to more appearance-related comments and likes, but also increased concern expressed by the community.
    \item We call for education strategies and policy regulations to better protect children's privacy and dignity.
\end{itemize}


\section{Children on TikTok}

\subsection{\emph{Sharenting}}
"Sharenting," that is, "parents sharing" information about their children on social media, has become a frequent phenomenon on social media \citep{amon2022sharenting,cataldo2022cradle,yegen2021sharenting}. While sharenting helps families stay connected and share joyful moments, it raises concerns about minors' privacy, and consent \citep{stephenson2024sharenting, walrave2023mindful}. 
The overall extent of sharenting is unknown and varies by country and platform. A survey of 493 US parents who regularly use social media found that nearly 90\% have shared content about their children online \citep{amon2022sharenting}. Similarly, in a survey of 2,900 Spanish children aged 9 to 17, 20\% reported that their parents shared information about them online, with older children and girls being more frequently affected \citep{garmendia2022sharenting}. 

Sharenting imperils minors' right to privacy, especially as few parents seem to ask their children for permission to disclose information  \citep{ni2022privacy}. In a survey with 1,460 Czech and Spanish parents, of whom around 80\% published pictures of their child, only 20\% obtained their child's consent \citep{kopecky2020phenomenon}. Some parents even deliberately ignore the will of their child, as reported in interviews with 12-14-year-olds \citep{ouvrein2019sharenting}. In the study by \citet{garmendia2022sharenting}, around 4\% of children reported negative outcomes from sharenting, such as hurtful comments, and 12\% of children requested their parents to remove shared content about them. Such information can include sensitive details: \citet{brosch2016child} revealed that among 168 parents' Facebook posts, 90.5\% mentioned their child's first name, 83.9\% shared birthdates, and 32.7\% uploaded personal documents or videos of the child.  

The content shared by parents often violates the child's dignity. \citet{stormer2023caregiver} identified TikTok videos containing psychological maltreatment towards children through caregivers, such as yelling at and pranking them. These videos received higher engagement in the form of likes, views, and comments than those without maltreatment of children. In an investigation by \citet{brosch2016child}, about 45\% of the parents posted photos that could be considered inappropriate, such as images of babies and toddlers in the nude or semi-nude, typically taken during baths or beach visits. Similarly, \citet{kopecky2020phenomenon} found that 20\% of the parents admitted having posted photos in which their children were partially exposed, and 3.5\% had shared photographs of their naked child at a neonatal or infant stage online. 

\subsection{Sexualization of Children on TikTok}
Previous research has linked skin exposure on social media to increased user engagement. On Instagram, for example, more revealing photos tend to attract more likes \citep{park2017private}. Additionally, a study on young women found that although self-sexualization rates in photos were relatively low, sexualized images garnered more likes and followers \citep{ramsey2018picture}. Further, non-government organizations have warned that algorithms may prioritize images showing more skin \citep{kayserbril2020undress}. However, this trend has not been specifically validated for TikTok or for content involving children.

While inappropriate content featuring children seems to be prevalent on many platforms, TikTok has been consistently criticized for enabling the sexual exploitation of children and adolescents \citep{polito2022compliance}. As the general level of sexualized behavior is high among the TikTok community, such behavior is likely to be imitated by young users, performing sensual or provocative dances or showing themselves in swimsuits or underwear \citep{suarez2023sexualising}. Additionally, children and adolescents receive sexually explicit comments and requests, as both interviews with minors \citep{soriano2023tiktok} and a BBC investigation of TikTok videos have shown \citep{silva2019}. Comments often focus on the physical appearance of children, complimenting their looks, and sometimes extend to inappropriate interactions, such as invitations to meet up, highlighting a dangerous aspect of online behavior towards minors \citep{silva2019}. Even more, a Forbes investigation has revealed child sexual abuse material being shared within private TikTok accounts \citep{levine2022}. 


Most investigations into online exploitation of children have been carried out by investigative journalists from major newspapers \citep{nyt2024,silva2019,levine2022,barry2021}, with scientific research on the subject being scarce. Although more comprehensive analyses of children on TikTok are emerging \citep{stephenson2024sharenting}, the majority of existing academic work in this area primarily focuses on qualitative reports and case studies \citep{khan2022new,soriano2023tiktok}. Large-scale quantitative analyses—essential for understanding the scale and patterns of such issues—remain scarce, demonstrating the need for further research on this topic.

\subsection{Child Protection Mechanisms on Social Media}
TikTok's guidelines prohibit harassment of minors through public or private interactions and commit to reporting content that endangers children to law enforcement \citep{tiktok2024}. Despite removing most sexually explicit comments within 24 hours of reporting, TikTok has not consistently eliminated messages that are inappropriate for children \citep{silva2019}.
At the same time, TikTok has increased efforts in content moderation of sexually explicit language on their platform, e.g., deleting videos that contain captions such as "sex". However, these automated detection algorithms can be circumvented by using alternative words and negative implications of the automated deletion of sexual content that might not be harmful but is aimed at educating young users \citep{steen2023you}. 

From a legal perspective, TikTok has been critiqued for not aptly protecting children's privacy. For example, researchers have found that TikTok enhances children's privacy protections in response to public outrage and regulatory pressures, not proactively as recommended by privacy frameworks \citep{polito2022compliance} or prioritizing profit over child protection measures \citep{salter2021need}.

\subsection{Scope and Research Questions}
This study addresses the critical issue of children's sexual exploitation and exposure on TikTok, emphasizing the need for targeted strategies and policies to protect young users from harmful content and interactions. While children's encounters with sexual content on platforms like TikTok are well-documented \citep{barry2021}, this study specifically assesses the risks associated with such exposures, rather than focusing on the broader issue of inappropriate content during online activities.
Since both effective mechanisms and comprehensive academic work in this area are lacking, we present a comprehensive study to assess the extent of children's exposure to different kinds on TikTok following these research questions: 

\begin{itemize}
    \item \textbf{RQ1}: How are children portrayed on TikTok?
    \item \textbf{RQ2}: How do users react to videos of children?
    \item \textbf{RQ3}: Can TikTok content featuring children be further traced to private devices and other websites?
    \item \textbf{RQ4}: Is there a relationship between the video content (i.e., the extent to which children are exposed or certain activities that are being performed) and user reactions? 
\end{itemize}

\section{Methods}

\subsection{Data}
Since the TikTok user guidelines do not permit users under 13 to hold accounts \citet{tiktok2024}, this study does not focus on accounts operated by minors. Instead, we examine TikTok accounts that feature children under 13 years old but are managed by adults, typically their parents.
To create our dataset, we searched for accounts showing children by using keywords, such as "child" or "kid". In addition, we included keywords suggested by the TikTok search console (e.g., "family," "child model"), in a snowball-like approach to also include accounts regularly active on TikTok.

For each matching account, we collected the IDs of the 100 first videos and excluded all videos that would not display a child below the age of 13 (e.g., because it either showed the parents or older siblings). To determine whether a child would fall below the age limit of 13, we used visual signs when clear (e.g., for toddlers) and age information given in the TikTok videos or profile description when unclear. We removed two accounts for which we could not determine the child's age in all videos based on our data. For each video, we collected the first 500 comments along with the video metadata (e.g., number of downloads). 

We extracted English-language comments containing text (beyond just numbers or emojis) from TikTok metadata to ensure consistency and reliability in our analysis. By focusing on English, the most widely used language on TikTok, we enhance the generalizability of our findings and minimize the risk of misinterpretation or context loss that can occur with multi-language analysis.
Our final dataset consists of 432,178 comments resulting from 5,896 unique videos of 115 TikTok accounts. 

\begin{figure}[h!]
  \centering
  \includegraphics[width=0.95\linewidth]{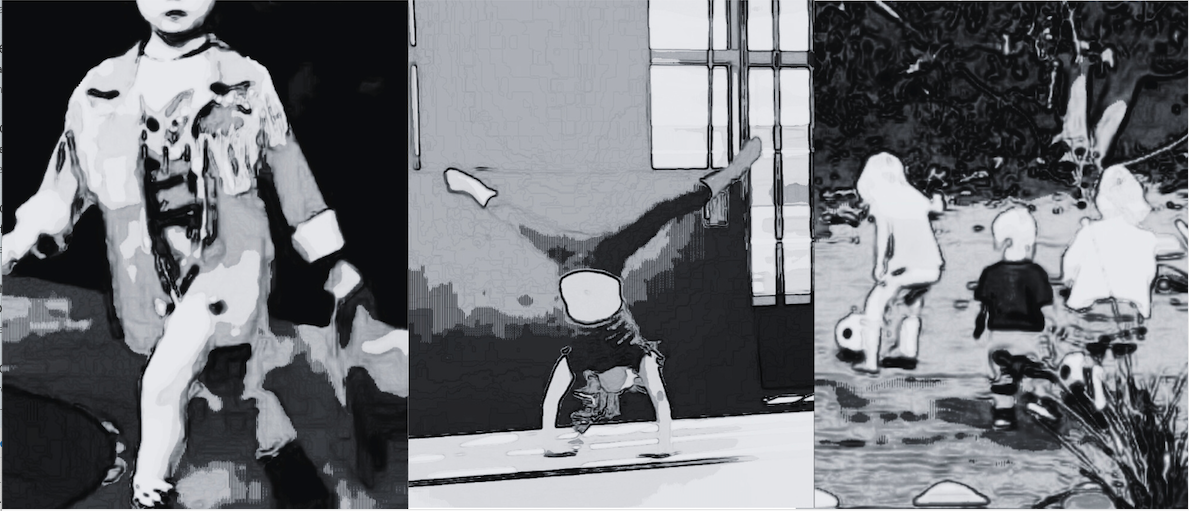}
  \caption{Example preview of child-related video content from each category (left to right: Fashion, Sports, Family).}
  \label{fig:preview}
\end{figure}

\subsection{Recurrent Themes}

Literature on concurrent TikTok themes indicates that a majority of content revolves around comedy, sports and fitness, beauty, and popular TikTok dances and challenges \citep{pryde2022tiktok, shutsko2020user, vaterlaus2021tiktok}. To answer RQ1 ("How are children portrayed on Social Media"), we analyzed the videos exploratorily, collecting frequently occurring themes, as well as noteworthy trends. We identified recurring themes consistent with previous research, such as beauty and sports \citep{pryde2022tiktok}. In addition, we found a large number of videos focused on day-to-day family life and introduced a third category, "Family". This led to the final categorization of accounts into the following groups (see Figure \ref{fig:preview} and Table \ref{tab:video_account_statistics}):

\begin{itemize}
    \item \textbf{Family:} content predominantly revolving around family-oriented themes on TikTok, including videos of parents, children, or both. This encompasses parents showcasing their daily routines with their children and performing TikTok dances or sketches together.
    \item \textbf{Sports:} content of children engaged in sports, predominantly gymnastics, and dancing. These accounts mostly show videos of the child or a group of children. 
    \item \textbf{Fashion:} fashion-related clips, which in most cases show one child in front of the camera, being put in various outfits to showcase different clothing styles and incorporating modeling poses.
\end{itemize}

\begin{table}[ht]
\centering
\small
\begin{tabularx}{\columnwidth}{Xrrr}
\midrule
\textbf{Category} & \textbf{Accounts} & \textbf{Videos} & \textbf{Comments} \\ 
\midrule
Family & 78 (67.8\%) & 4,073 (69.1\%) & 340,921 (78.9\%)\\
Fashion & 21 (18.3\%) & 1,336 (22.6\%) & 83,708 (19.4\%)\\
Sports & 16 (13.9\%) & 487 (8.3\%) & 7,549 (1.7\%)\\
\midrule
Total & 115 & 5,896 & 432,178 \\ \midrule
\end{tabularx}
\caption{Number of accounts, videos, and comments per account category.}
\label{tab:video_account_statistics}
\end{table}

\subsection{Video analysis}

\subsubsection{Video Annotation}
After identifying frequently occurring categories within our dataset, we manually annotated each video (\textit{n} = 5,896) with the following attributes, referring to the child(ren) in the focus of the video: First, we annotated the child's perceived gender\footnote{We acknowledge and respect that gender identity is diverse and can extend beyond traditional male and female categories. Our use of visual cues or descriptions to categorize gender is based on conventional perceptions and does not intend to exclude or invalidate non-binary, genderqueer, or other gender identities.} to account for potential differences, categorizing each instance as female, male, or both (when multiple children of different genders are featured). When we could not detect a gender based on the video, the gender was categorized as 'unknown'. 
Second, we analyzed the level of skin exposure in each video, consistent with our research question. We marked videos as showing skin exposure if the child appeared in revealing clothing, i.e., being naked, in swimwear, or wearing outfits exposing the belly, or entire the upper or lower body.
Third, we annotated whether a child was shown wearing makeup to account for additional appearance-based factors, as makeup can significantly alter how a child is perceived. This was marked as true if the child was wearing clearly visible makeup.

One of the authors of this study and a research assistant conducted the labeling of the 5,896 videos included in our dataset. Both researchers labeled the full dataset individually. We calculated Cohen's Kappa for inter-annotator agreement \citep{cohen1968weighted}, yielding a score of $\kappa=.67$ for skin exposure, indicating substantial agreement, and $\kappa=.41$ for the makeup category, indicating moderate agreement. Ambiguous cases were discussed within the research team, and a joint label was decided upon collectively.
The moderate agreement for the makeup category reflects the challenge of distinguishing between actual makeup and visual filters commonly applied to videos, which often enhance the appearance of makeup.

\subsubsection{Comment Classification}
\label{sec:meth_comm}
We then studied user reactions towards videos depicting children (RQ2) by evaluating the video comments. We used quantitative methods to get a first overview of frequent reactions, analyzing the most frequent words, bigrams, trigrams, and emojis in user comments. 

During this analysis, it became evident that two categories were exceedingly prevalent within the dataset: (1) comments on the \textbf{visual appearance} of the child (\textit{"She is so pretty,"} \textit{"He has beautiful eyes"}) and (2) comments showing a very strong \textbf{affection} towards the child (e.g., \textit{"I love you,"} \textit{"You are my girl"}). We created two dictionaries for extracting these comments and, given the high number of results, analyzed them using quantitative statistics.\footnote{Code and dictionaries are available at: \url{https://osf.io/huf76/?view_only=4dbfb7991f3e47b0af2cb07b2cad6c45}}

Next, we investigated whether children on TikTok were targeted by \textbf{inappropriate comments} (referring to the child as a sex object), whether these accounts received \textbf{contact offers}, and whether other users expressed \textbf{concerns} about the child’s exposure on the platform. We developed three dictionaries, selecting keywords, bigrams, trigrams, and emojis that were most likely to indicate these categories. The selection process involved thorough consideration to ensure the terms would effectively capture the nuances of inappropriate behavior, contact attempts, and concerns (e.g., words like "send" and "address" for contact attempts). This approach resulted in a set of 100,043 comments featuring inappropriate comments, contact offers, or concerns. We inspected each of these comments manually and labeled them with one of the codes: inappropriate, contact, concern, or none. For unclear comment meanings, we rewatched the corresponding video to ensure correct classification.

While contact offers and concern about exposure were relatively easy to detect, recognizing inappropriate comments was more challenging: As many videos showed the parent and the child, there were numerous sexually explicit comments for which it was unclear to whom they were addressed (e.g., \textit{"sexy"}), and which we thus did not classify as inappropriate. Additionally, recognizing the subjectivity of appropriateness across different cultures, many comments that initially appeared inappropriate for children were reconsidered. This led us to adopt a conservative stance in our classification, opting not to label comments as inappropriate unless they were unequivocally so in a broad cultural context, focusing on the most overt instances of inappropriateness.

Regarding RQ3, we studied whether and to which extent the content shown on TikTok was distributed a) on private devices and b) on websites other than TikTok (RQ3). To answer the first part of the question, we inspected the number of times users downloaded a video. For the second part, which was mainly motivated by one comment in our data, expressing that the TikTok video was used on a child pornography website, we employed two strategies: First, for each account, we used the Bing reverse image search utilizing a screenshot of the child of one of the videos. Second, we searched for the username of each account using Bing image search. We then evaluated the search results and collected all websites that used a copy of an image or video of the child from TikTok.

Finally, to answer RQ4, we conducted a quantitative analysis to compare characteristics between two groups of videos: those displaying children wearing exposing clothing and those that do not. We employed statistical tests to examine differences in various metrics, including attachment, appearance-based comments, offers of contact, concerns raised by other users, and engagement metrics, such as the number of likes and downloads. Means and standard deviations were calculated for each variable, and \textit{t}-tests were applied to determine statistical significance between groups. We applied Bonferroni correction to account risk of Type I errors due to multiple comparisons.

\subsubsection{Topic Modeling}
For further content evaluation, we applied BERTopic \citep{grootendorst2022bertopic} to identify common topics. This topic modeling technique leverages BERT embeddings and Term Frequency-Inverse Document Frequency (TF-IDF) to cluster semantically similar comments, providing a more nuanced understanding compared to traditional methods like Latent Dirichlet Allocation (LDA). Each preprocessed comment was transformed into a vector representation using pre-trained BERT embeddings. 
To simplify and visualize the data, BERTopic reduces its dimensionality using UMAP (Uniform Manifold Approximation and Projection) and applies clustering through HDBSCAN (Hierarchical Density-Based Clustering). 
We specified the number of topics ($\textit{k} = 50$) to strike a balance between capturing detailed nuances and maintaining broader thematic coherence. The resulting topics were thoroughly analyzed, with their top words inspected to ensure they were meaningful. We excluded topics predominantly composed of (account) names, as these do not provide significant insight into the content. Finally, we visualized the topics using an intertopic distance map, which displays the relationships and similarities between the identified topics, enhancing our understanding of the data's thematic structure.

\section{Results}

\begin{table*}[ht!]
\centering
\begin{tabular}{llcccc}
\toprule
\textbf{Category} & \textbf{Gender} & \textbf{Exposed} & \textbf{\% Exposed} & \textbf{Makeup} & \textbf{\% Makeup} \\ 
\midrule
\multirow{3}{*}{Family (\textit{n} = 4,073)} & Female & 372 & 9.13\% & 13 & 0.32\% \\ 
& Male & 97 & 2.38\% & 0 & 0\% \\ 
& Both & 106 & 2.60\% & 3 & 0.07\% \\ 
\textbf{Total} && \textbf{584} & \textbf{14.34\%} & \textbf{16} & \textbf{0.39\%} \\ 
\midrule
\multirow{3}{*}{Fashion (\textit{n} = 1,336)} & Female & 327 & 24.48\% & 176 & 13.17\% \\ 
& Male & 5 & 0.37\% & 9 & 0.67\% \\ 
& Both & 12 & 0.90\% & 9 & 0.67\% \\ 
\textbf{Total} && \textbf{344} & \textbf{25.75\%} & \textbf{194} & \textbf{14.52\%} \\ 
\midrule
\multirow{3}{*}{Sports (\textit{n} = 487)} & Female & 183 & 37.58\% & 9 & 1.85\% \\ 
& Male & 20 & 4.11\% & 0 & 0\% \\ 
& Both & 23 & 4.72\% & 1 & 0.21\% \\ 
\textbf{Total} && \textbf{226} & \textbf{46.41\%} & \textbf{10} & \textbf{2.05\%} \\ 
\midrule
\multirow{4}{*}{\textbf{Overall (\textit{n} = 5,896)}} & Female & 882 & 14.96\% & 198 & 3.36\% \\ 
& Male & 122 & 2.07\% & 9 & 0.15\% \\ 
& Both & 141 & 2.39\% & 13 & 0.22\% \\ 
& Unknown & 9 & 0.15\% & - & 0\% \\ 
\textbf{Total} && \textbf{1,154} & \textbf{19.57\%} & \textbf{220} & \textbf{3.73\%} \\ 
\bottomrule
\end{tabular}
\caption{Video content by category and gender. Percentages for Exposed and Makeup content are calculated relative to the total number of videos in that specific category (e.g., the first row presenting absolute numbers and the share of videos featuring exposure of children within the Family category).}
\label{tab:all-overview}
\end{table*}

Across the 5,896 videos collected and analyzed, 19.57\% show children in exposed clothing, and 3.73\% show children wearing makeup (Table \ref{tab:all-overview} and \ref{tab:gender-variables}). 
Exposed clothing is highest within the Sports category, appearing in almost half of the videos (46.41\%), which could be related to the nature of sports content, such as minors wearing revealing gymnastics attire. In contrast, makeup is present in only 2.05\% of sports videos. 
Makeup is most prominent in the Fashion category (14.52\%), where exposed clothing is observed in 25.75\% of the videos, suggesting a potential association with fashion-related content. 
The Family category shows the least number of children wearing makeup (0.39\%), with 14.35\% of videos depicting skin exposure.


Looking at gender differences, we find that videos featuring female children are the most prevalent, comprising 66.79\% of the total dataset, and girls are more often presented in exposed clothing (14.96\%) and wearing makeup (3.36\%) than boys (2.07\% exposed, 0.15\% makeup).
Videos with children of both genders constitute 13.52\% of the total, with 2.39\% showing exposure and 0.22\% depicting makeup use (Table \ref{tab:gender-variables}). These findings underscore a significant gender disparity, with girls being exposed more frequently, highlighting concerns regarding gendered presentation and objectification in social media content.

\subsection{Comment Analysis}

\subsubsection{Comments on Visual Appearance}
\begin{figure}[h!]
  \centering
  \fbox{
    \includegraphics[width=0.85\linewidth]{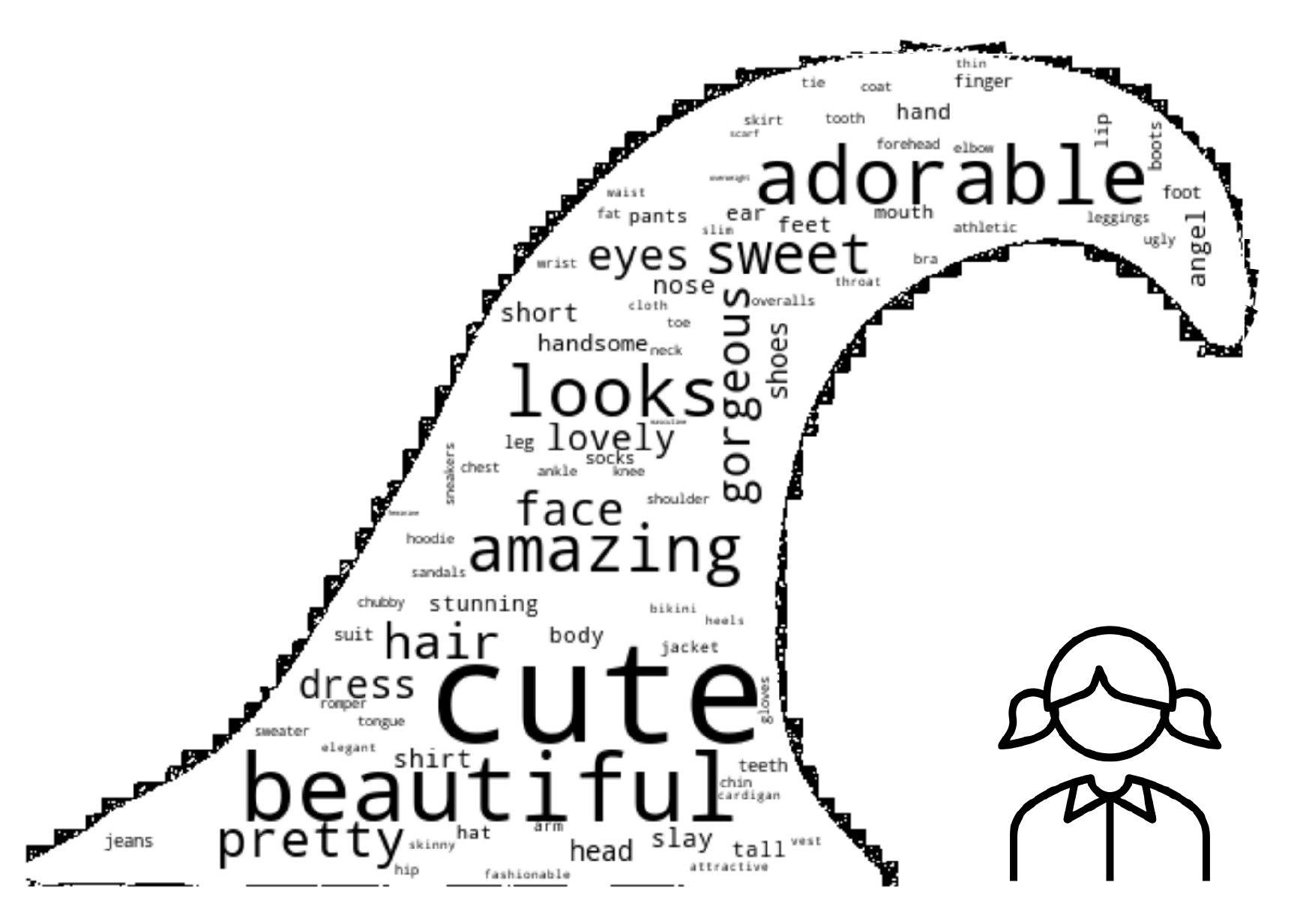}}
  \caption{Overview of most common appearance-related words in our dataset.}
  \label{fig:wave}
\end{figure}

\begin{figure*}[t!]
  \centering
  \includegraphics[width=\textwidth]{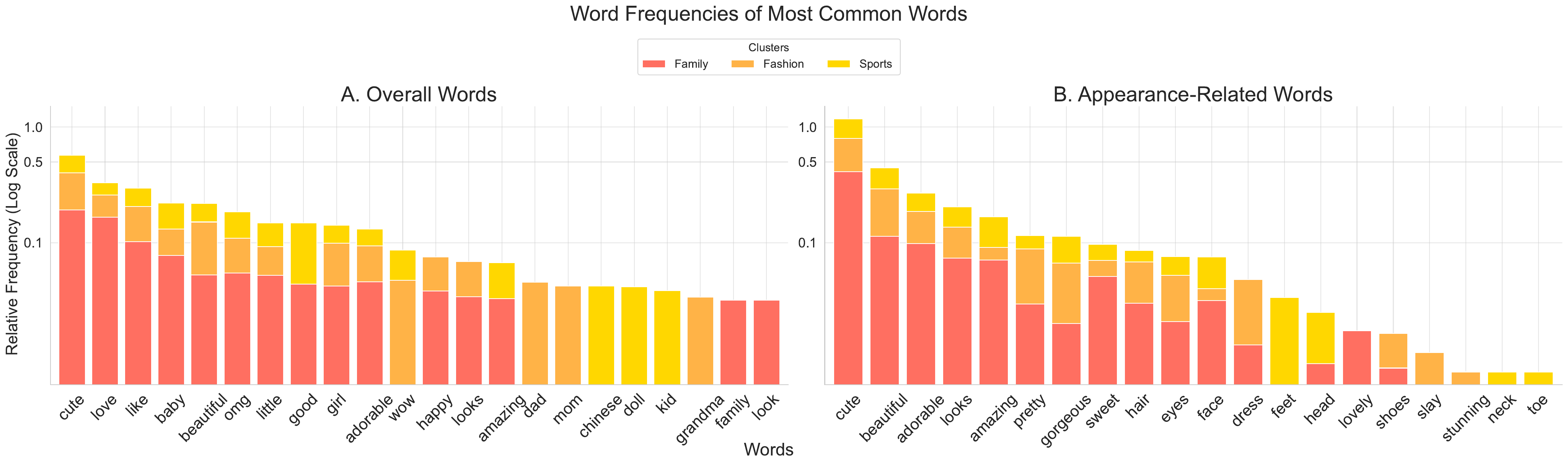}
  \caption{Most frequent appearance-based words per category.}
  \label{fig:word-frequencies-app}
\end{figure*}

In 88,627 comments (79.0\% to Family accounts, 19.32\% Fashion, 1.78\% Sports), words related to visual appearance were used. Under the most prominent ones were cute, beautiful, adorable, amazing, sweet, pretty, and gorgeous, as well as hair, face, eyes, and dress (Figure \ref{fig:wave} and \ref{fig:word-frequencies-app}). Although most of these comments were written in a very positive tone, there were also some negative comments (e.g., \textit{"y does China want me watching ugly middle eastern children"} or \textit{"I hope these two [children] improve with time. Looks are pretty disappointing watching they have really attractive parents."})

When analyzing the most frequent words, we found a substantial overlap between appearance-based words and overall most frequent words (Figure \ref{fig:word-frequencies-app}). This was particularly true for general terms, such as \textit{"cute"}, \textit{"beautiful"}, etc. Inspecting only appearance-based words, we see that body parts, such as \textit{"face,"} \textit{"hair,"} and \textit{"eyes,"} belonged to the most frequent words in all categories. For the Fashion and Family category, we observed \textit{"dress"} and \textit{"shoes"} as the most frequent terms relating to what the children are wearing. On the contrary, the 15 most frequent words in the Sports category did not feature any clothing but mentioned body parts more frequently, such as \textit{"feet,"} \textit{"neck,"} and \textit{"toe"}. This shift suggests a focus on physical performance and anatomical aspects rather than attire. 
In the Fashion category, further words such as \textit{"stunning"} and \textit{"slay"} appeared but were not present in the other categories, indicating a distinct emphasis on style and presentation not observed in other categories.

\begin{table}[ht]
\centering
\small
\begin{tabular}{llll}
\toprule
\textbf{Gender} & \textbf{Exposed} & \textbf{Makeup} & \textbf{Total} \\ \midrule
Female & 882 (14.96\%) & 198 (3.36\%) & 3,938 (66.79\%) \\ 
Male & 122 (2.07\%) & 9 (0.15\%) & 1,083 (18.37\%) \\ 
Both & 141 (2.39\%) & 13 (0.22\%) & 797 (13.52\%) \\ 
Unknown & 9 (0.15\%) & - & 78 (1.32\%) \\ \midrule
\textbf{Total} & 1,154 (19.57\%) & 220 (3.73\%) & 5,896 (100\%) \\ 
\bottomrule
\end{tabular}
\caption{Overview of gender distribution. Percentages indicate the proportion of videos for the label relative to the total number of videos for the respective gender.}
\label{tab:gender-variables}
\end{table}

\subsubsection{Inappropriate Comments, Attachment \& Contact Offers}
We found 12 clearly inappropriate comments directed toward children in the dataset. While 12 may seem like a relatively small number, the nature of these comments still makes them concerning. These comments were all directed at videos from eight unique accounts (four Fashion and four Family accounts), with six of the videos featuring toddlers and two showing around 6-year-olds. These comments included content such as \textit{"sexy bi**h,"} \textit{"I like the way you suck on your glasses,"} or \textit{"Hot babies"} referring to toddlers performing model poses in front of the camera, \textit{"Save her for me when she's 18+"} referring to a toddler in pajamas, or \textit{"That laugh at nothing makes me want to kiss you with a lot of passion and marry you [...]"} referring to a six-year-old playing dolls with her dad. Five of these comments referred to videos showing children in exposing clothing. 
The presence of any such comments is significant and troubling, especially given the vulnerable age of the children involved.

Moreover, a significant number of strangers on TikTok expressed some forms of \textbf{attachment} to the depicted children. In 4,206 comments, users reported their love ("love you/him/her," "love this/your baby/girl/boy"), and in 122 comments users referred to the child as "my girl/boy/baby" (e.g., \textit{"my baby girlfriend,"} \textit{"[...] I love you my baby doll you are so sweet,"} \textit{"[...] dance for me my girls [heart-emojis]"}). In numerous comments, users asked to adopt the child, with some expressing more than joking intentions (e.g., \textit{"she's just so cute, I love [heart-emoji] her. can I adopt her \& I'm serious. I lost my baby when I was in a coma (pregnant)"}). In addition, multiple comments showed a strong protection motive towards the child, even though there was no need for protection expressed in the respective video (\textit{"I will protect this child with my life. she too precious and seems so sweet",} \textit{"I'm ready to donate myself to protect that angel [...]"}).

While expressions of affection are common on TikTok, the strength of bonding some users seemed to establish towards the child should not be underestimated. In our dataset alone (which only included comments of at most 100 videos per account), we found 150 users who sent more than 30 comments to a single account, and it was unclear whether these users were strangers or acquaintances to the child. For example, one user with a private account sent 82 comments to a toddler, such as \textit{"Hello little love!!,"} \textit{"Sweet [Child's name]!!,"} or \textit{"Hi cutie pie [Child's name]!!"}. Another user sent 79 comments to a toddler's model account ranging from \textit{"She so beautiful,"} \textit{"Awwww you are a pretty girl,"} and \textit{"I can watch she over n over [...]."}

In an additional 114 comments, users tried to contact the account holder, e.g., by asking to exchange direct messages or addresses. While many offers included collaboration requests, possibly due to an influencer role of the account, it was not possible to determine whether real companies, content creators, or private interests stood behind these offers. Also, multiple users asked for an address to send gifts to the child, but it was unclear whether this was due to company interests in content creators, admiration of fans, or darker intent. Further, there were several comments ranging from nice and joking messages to disturbing ones, depending on the interpretation, e.g., \textit{"I would pay you to babysit her,"} \textit{"mami plz can u send me pic in may tik tok acount [...] am in huge luv wz ur baby boyyy [...]"} or \textit{"[...] is it possible for you to send me a box of all her old dresses [...]"}.

\subsubsection{Comments Expressing Concerns}

In 560 comments, users expressed concerns about the way the child was depicted in the video. This included worries about clothing, e.g., \textit{"You need to learn how to dress your child because that is so inappropriate"}, and the context in which the child was shown (e.g., suggestive poses, performance of 18+ related TikTok trends). Users also raised concern about the number of downloads (e.g., \textit{"It's terrifying how many saves this has. Do you think a bunch of random well-intended people save these videos?,"}) the form of comments (e.g., \textit{"There are some disgusting comments on this video. Please if this is really your daughter, protect her from grown men who are watching these,"}) or the future of the child (\textit{"I can't even imagine how she will feel in 10 years knowing millions of people saw this. It's so sad you choose money over her well being and privacy"}). Some even criticized TikTok's regulations concerning such content, e.g., \textit{"I really wish TikTok would ban minors from being in videos [...]"} or \textit{"how is this not inappropriate??? and why can't I block this app from showing me this??? [...]"}.
Interestingly, there seemed to be a strong common ground in which videos/accounts were regarded as inappropriate by the community: 206 (36\%) of concerned comments were directed to a fashion account of an around six-year-old girl and 153 (27\%) to a fashion account of an around three-year-old girl. Further four fashion and three family accounts received 10-40 comments (together 28\%), and 88 accounts in our dataset received no such comment at all.

\subsection{Exposure and Its Influence on Viewer Interaction}

The results from the \textit{t}-tests offer comparisons between videos with exposure and those without (Table \ref{tab:group differences}). Videos with exposure received the a similar percentage of attachment comments than those without (\textit{t} = 1.71, \textit{p} = .09) as well as a similar share of contact offers (\textit{t} = .47, \textit{p} = .64). Conversely, videos with exposure had a higher share of appearance-based comments (\textit{M} = .2360, SD = .42) than those without (\textit{M} = .1847, \textit{SD} = .39, \textit{t} = -29.63, \textit{p} $< .001$). Raised concerns revealed a significant discrepancy, with videos with exposure (\textit{M} = .0037, \textit{SD} = .06) exhibiting more raised concerns in comments than those without exposure (\textit{M} = .0009, \textit{SD} = .03, \textit{t} = -17.86, \textit{p} $< .001$). This highlights the impact of exposure on both comments related to appearance and the level of concern expressed by viewers.

The number of likes and downloads showed significant differences between videos with and without exposure. Videos with exposure received more likes on average (\textit{M} = 413,704.20, \textit{SD} = 931,821.27) than those without exposure (\textit{M} = 387,034.25, \textit{SD} = 903,999.81,\textit{t} = -6.66, \textit{p} $< .001$). At the same time, videos with exposure had fewer downloads on average (\textit{M} = 3,203.55, \textit{SD} = 9,511.63) than to those without exposure (\textit{M} = 5,857.85, \textit{SD} = 39,301.72, \textit{t} = 16.42, \textit{p} $< .001$). In summary, while exposure appears to increase the likelihood of receiving likes, it inversely correlates with the number of downloads, highlighting the complex dynamics of audience engagement in digital environments.

\subsection{Secondary Distribution of Child-Related Content}
On TikTok, users can download videos, with the metadata revealing the number of downloads. First, we examined how frequently users saved videos of children to their own devices, which complicates efforts by platforms and parents to remove such content later. Each video was downloaded on average 137 times, while this number heavily varied per video. While 35\% of videos were not downloaded at all, 43\% were saved between 1 and 100 times, 13\% between 100 and 1,000 times, and 8\% more than 1,000 times, with a video of a sneezing baby receiving the maximum number of 1,069,362 downloads. Download numbers were significantly positively correlated to the popularity of the video, measured in the number of views, likes, comments, and shares (all Pearson correlations with p $< .001$). 

In a second step, we investigated whether users would share or repurpose these videos or parts of them on other platforms, potentially without the permission of the parental guardians. For 23 accounts, we found copies of the content on other platforms and web pages. Besides various other social media platforms, like Instagram or Facebook, these websites included a platform containing duplicates of all TikTok videos with the defined goal to let users watch TikTok videos in an anonymous way; numerous Pinterest collections with pictures of children, some ordered by child and name, others tagged or titled with words such as \textit{"cute babies"} or \textit{"[...] dancing like a stripper"}; and multiple websites on 'Social media celebrities', containing profiles of numerous children on TikTok, including information on the child's full name, birthplace, birth date, height, waist and dress size, medical information, and more.

\begin{table*}[h]
\centering
\begin{tabular}{lcccc}
\toprule
 & \textbf{Videos with Exposure} & \textbf{Videos without Exposure} & & \\ 
 & (\textit{n} = 1,154) & (\textit{n} = 4,742) & & \\ 
\midrule
 & \textit{M (SD)} & \textit{M (SD)} & \textit{t} & \textit{p} \\ 
\midrule
Attachment & .0094 (.10) & .0101 (.10) & 1.71 & 0.09 \\
Appearance & .2360 (.42) & .1847 (.39) & -29.63 & $< .001$*** \\
Contact Offers & .0002 (.02) & .0003 (0.02) & .47 & .64 \\
Expressed Concerns & .0037 (.06) & .0009 (.03) & -17.86 & $< .001$*** \\
N Likes & 413,704.20 (931,821.27) & 387,034.25 (903,999.81) & -6.66 & $< .001$*** \\
N Downloads & 3,203.55 (9,511.63) & 5,857.85 (3,9301.72) & 16.42 & $< .001$*** \\
\bottomrule
\end{tabular}
\caption{Group comparisons for videos with content of exposed children and without. Numbers indicate the share of comments featuring attachment, appearance, contact offers, and expressed concerns, alongside absolute numbers for likes and downloads, *** indicating a \textit{p}-value below .001.}
\label{tab:group differences}
\end{table*}

\subsection{General Video Content}
The videos in our data span a wide range of content, from everyday life and sports to family performances and modeling. These videos feature children across various age groups, with a significant focus on pregnancy, childbirth, and early child-rearing. Content often includes intimate aspects of family life, such as prenatal appointments, childbirth experiences, and newborn care, with creators sharing advice on routines, breastfeeding, and sleep schedules.

"Routine videos" are common, showcasing day-to-day activities with toddlers, from morning preparations to bedtime rituals. These videos exposed detailed aspects of personal lives to a large audience, raising privacy concerns.

Certain trends raise further concerns, especially those emphasizing children's physical appearance. Some videos featured parents criticizing their child's appearance or making comparisons between themselves and their children. One trend involved parents posting a photo with the caption \textit{"When you think I'm pretty..."} followed by an image of their child with \textit{"... you should see my daughter(s)."}
Additionally, there were instances where children appeared in inappropriate contexts, dancing to mature songs or participating in age-inappropriate TikTok challenges. Some parents also mixed suggestive adult content with footage of their children, raising concerns about the appropriateness of such content.

Figure \ref{fig:word-frequencies-app}A showcases the top 15 most common words in the three distinct categories Fashion, Family, and Sports, highlighting both overlaps and differences in word usage across these categories.
Words such as \textit{"cute,"} \textit{"like,"} \textit{"love,"} \textit{"baby,"} \textit{"omg,"} \textit{"beautiful,"} \textit{"little,"} \textit{"one,"} and \textit{"girl"} appear across all three clusters, suggesting universal themes of affection, admiration, and personal interest that transcend specific contexts. 
Comments on videos in the fashion category are characterized by a blend of aesthetic appreciation (\textit{"beautiful,"} \textit{"adorable"}) and social/family roles (\textit{"dad,"} \textit{"mom"}). The consistent presence of family-related terms alongside fashion-centric vocabulary suggests a notable connection between familial themes and modeling. For instance, discussions might revolve around scenarios where children are accompanied by their parents during fashion events, or there may be references to familial pride, such as mentions of a \textit{"proud mom"} supporting her child's endeavors in the fashion world. 
The family category shows a stronger emphasis on personal and relational expressions (\textit{"love,"} \textit{"baby,"} \textit{"little,"} and \textit{"adorable"}) alongside a higher frequency of words, indicating more intense discussions or more content volume around family topics.

\subsection{Compliments and Cuteness as Recurring Topics}

\begin{figure*}
    \centering
    \includegraphics[width=0.98\linewidth]{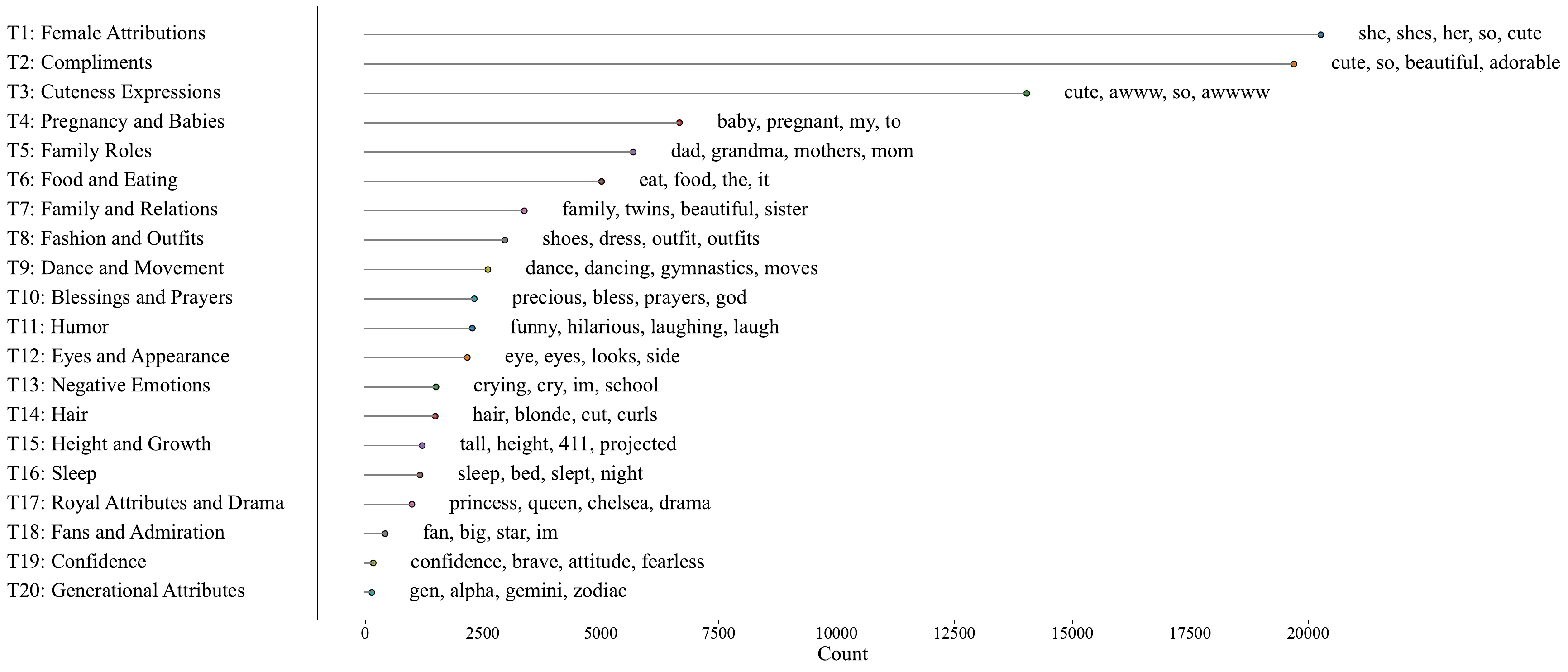}
    \caption{Overview of topics and their most salient words (selected topics based on their coherence).}
    \label{fig:topic-counts}
\end{figure*}

The results from BERTopic aligned with our initial observations about the video content.
The topics and their most significant words are detailed in Figure \ref{fig:topic-counts}. After excluding topics mainly composed of names (e.g., TikTokers or children) and those lacking coherent themes, we focused on 20 meaningful topics.\footnote{A full list of topics is shared with the code.} Key categories included descriptions of physical appearance, expressions of cuteness, family dynamics, and lifestyle interests such as food, fashion, and dance.
The dataset shows a significant variation in topic prevalence, with "Compliments" leading
with 20,269 comments, while topics like "Generational Attributes" have 143
comments. This distribution underscores the dominance of themes related to compliments,
cuteness, and family, with a strong emphasis on positive and affectionate language, as seen in "Compliments," "Cuteness Expressions," and "Humor," highlighting admiration and endearment throughout the dataset. Family and relational dynamics also feature prominently, as seen in topics like "Family Roles" (Topic 8), "Pregnancy and Babies" (Topic 6), and "Family and Relations" (Topic 17), suggesting a strong focus on familial relationships and life events. Additionally, topics such as "Fashion and Outfits" (Topic 19) and "Hair" (Topic 30) point to an interest in personal appearance and style.

\begin{figure}[h]
  \centering
  \includegraphics[width=0.98\linewidth]{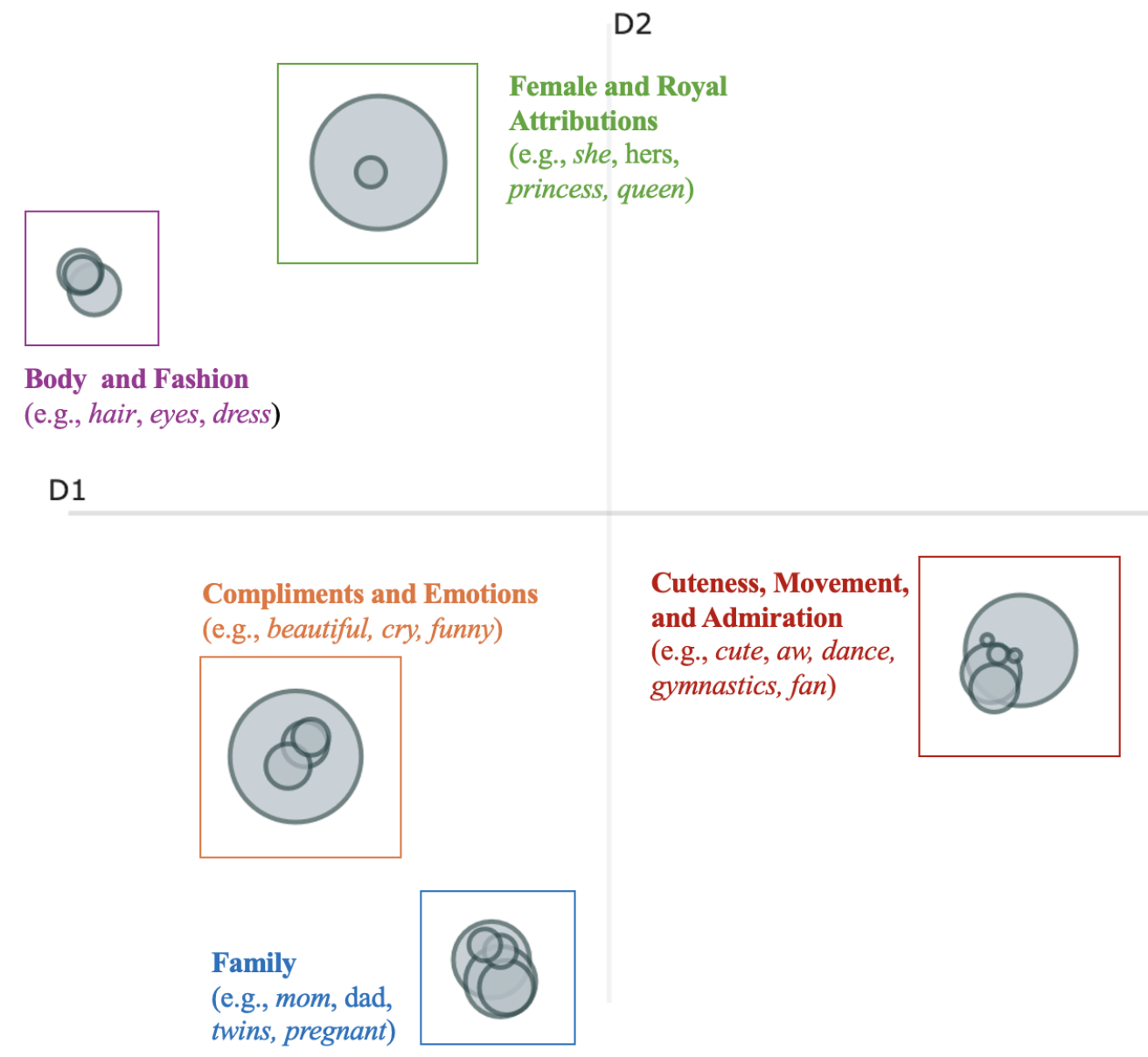}
  \caption{Intertopic Distance Map: Visualization of the relationships between topics, based on their embeddings, displayed in a 2D space.}
  \label{fig:topic-map}
\end{figure}

The intertopic distance map in Figure \ref{fig:topic-map} provides a visual overview of the relationships
between topics identified by the BERTopic model. 
The "Body and Fashion" cluster in the upper-left quadrant centers around physical appearance and style, featuring terms like \textit{"hair,"} \textit{"eyes,"} and \textit{"dress,"}.
The "Family" cluster positioned in the lower-middle emphasizes familial roles and relationships, highlighted by words such as \textit{"mom,"} \textit{"dad,"} \textit{"twins,"} and \textit{"pregnant."} Conversely, the "Cuteness, Movement, and Admiration" cluster in the lower-right underscores endearing qualities and activities with terms like \textit{"cute,"} \textit{"dance,"} \textit{"gymnastics,"}. The close proximity of circles within each cluster indicates strong thematic connections, while the distinct separation between clusters, such as between "Female and Royal Attributions" and "Body and Fashion," suggests diverse areas of discourse within the dataset.

\section{Discussion}

This paper presents one of the first studies to comprehensively examine the impact of exposure of children on TikTok on user engagement and interaction. Nearly 20\% of the analyzed videos feature at least one child in revealing clothing, representing a significant proportion. This finding aligns with previous research on the widespread presence of revealing images of minors across various platforms \citep{kopecky2020phenomenon}. Our analysis shows that users often react strongly to children online, not only through inappropriate comments or contact offers but also by expressing intense forms of attachment. We further found significant differences between videos featuring children in exposing attire and those without, notably in comments related to appearance and expressed concerns, with the former receiving more comments on appearance and concerns, as well as more likes but fewer downloads. 
This observation could suggest a discrepancy in engagement behavior among viewers. Viewers might appreciate the content enough to react positively with likes in the moment as a normal, short-term reaction, yet they may not find it compelling or relevant enough to save for later consumption. Another explanation is that viewers might be reluctant to have such content appear in their TikTok download history, leading them to use alternatives like screencasts.

Although \citet{stormer2023caregiver}'s study focused on child maltreatment rather than exposure, their analysis revealed similar patterns in user engagement with content involving children. This suggests that certain forms of inappropriate behavior could drive increased engagement. However, given the differing subjects, a direct comparison, particularly regarding likes and appearance-based comments, is challenging and calls for future research in this field.


Besides the high prevalence of sharenting on social media, our results also show that parents often face criticism from other users for sharing videos of their children online. This reaction underscores a growing awareness and concern about the potential dangers of exposing children on these platforms, including risks like cyberbullying, exploitation, and unwanted attention. Our findings align with previous research, which also emphasizes the complex social dynamics and challenges parents encounter when navigating public perceptions of sharenting \citep{nyt2024, stephenson2024sharenting}.

\subsection{Limitations and Future Research}

\subsubsection{Video Selection and Category Distribution}
We primarily utilized keyword-based search, likely influenced by automated recommendations. While this approach yielded a diverse range of TikTok accounts, it may not fully capture the entirety of content featuring children on TikTok. Additionally, while our sample represented various cultural backgrounds, it may not capture the full spectrum of underage users presented on TikTok across different regions and cultures. Future research should aim to explore alternative methodologies for accessing and analyzing content featuring minors on platforms like TikTok.

With the Family category comprising 67.8\% of accounts, 69.1\% of videos, and 78.9\% of comments, our analyses might be skewed towards family content. However, since this distribution reflects real-world trends where such content is highly prevalent and engaging on TikTok, our findings reflect typical user interactions.
Further research should aim to confirm the prominence of these categories while also exploring opportunities to create a more balanced dataset.

\subsubsection{Social Acceptance}
The social acceptance of children wearing revealing clothing or being depicted in minimal attire, such as diapers, on social media might vary with context and with age. For babies and very young children, posting images where they are naked or in diapers is often seen as more acceptable, reflecting societal norms that view such depictions as innocuous or adorable representations of early childhood. However, as children grow older, societal expectations and concerns about privacy and appropriateness come into play, leading to a decrease in the acceptance of sharing images that expose too much. For example, the share of 45\% of exposing baby pictures on Facebook in \citet{brosch2016child}'s study is substantially higher than what we found when looking at a diverse age group. Another example is sportswear: attire that is often short and reveals the midriff might be more accepted for children of various ages due to the specific context of athletic activities.
The discrepancy between parents posting videos of children in revealing attire and other users voicing concerns in the comments highlights a clear divergence in perspectives.
Our findings, showing that exposing videos tend to attract more concern, are consistent with research indicating a rising awareness on privacy issues with children's videos \citep{walrave2023mindful}. 

\subsubsection{Educational \& Policy Implementations}

There is a growing call from researchers for more focused studies to enhance parental awareness on sharenting \citep{barnes2020sharenting,williams2021combating}. Our findings suggest that many parents do not fully recognize the privacy implications of sharenting, underscoring the need for targeted educational interventions. Such interventions should aim at increasing parents' understanding of potential risks, knowledge of the platform's safety features, and general data literacy \citep{taylor2024parenting}. 
While there are many online resources to support parents, such as guides from the TikTok Safety Center, ConnectSafely \citep{connectsafely2023} and Internet Matters \citep{internetmatters2023}, there is a need for strategies to target parents with less motivation or more fear to engage with responsible social media use. Previous research has shown that parents engage in mindful sharenting primarily due to previous negative experiences and a desire to protect their child's privacy \citep{walrave2023mindful}.  
For future studies, it would be valuable to analyze how parents adapt and learn from online safety guidelines and whether there are noticeable changes in their behavior and attitudes before and after an educational intervention \citep{williams2021combating}.

Additionally, enhanced TikTok regulation is required to create a safer environment for children. Policymakers should improve age verification, reporting mechanisms, and transparency in content moderation. National and international measures are necessary to regulate the online presentation of minors and protect their privacy and dignity, even when it conflicts with parents' financial or social incentives.

\subsection{Conclusion}
Our research highlights the complex dynamics surrounding sharenting. It uncovers worrying trends in the presentation of minors online and the user reactions such content evokes. It reflects a growing societal concern about the exposure of children and the related risks. This study thus emphasizes the urgent need for effective strategies to protect young users, both from an educational and policy perspective.
\section{Ethics Statement}
This study involved several ethical considerations. While there is a general societal interest in understanding potential risks or harms to children on social media, these concerns may not always align with the interests of account creators, guardians, or TikTok itself. Thus, we had to carefully balance the need to minimize harm with the importance of informed consent. Given that the children's privacy had already been compromised due to widespread viewing, commenting, and downloading of their content, we aimed to avoid bringing additional attention to these children.
To adhere to ethical standards, we did not share any identifying details such as account names, pictures, video links, or non-aggregated metadata. Anonymized comments were included only after ensuring they could not be used to identify individuals or accounts through search engines or TikTok. Data will not be shared and is stored only for the purpose of analysis and according to the university's ethics guidelines. Although we were prepared to report any content classified as child pornography under COUNTRY ANONYMIZED law, we did not encounter such material. All data analyzed was from publicly accessible sources, and our study did not involve direct research with human subjects.


\bibliography{aaai25}

\end{document}